\documentclass{article}

\usepackage{arxiv}

\usepackage[utf8]{inputenc} 
\usepackage[T1]{fontenc}    
\usepackage{hyperref}       
\usepackage{url}            
\usepackage{booktabs}       
\usepackage{amsfonts}       
\usepackage{nicefrac}       
\usepackage{microtype}      
\usepackage{multirow}  
\usepackage{graphicx}
\usepackage{xcolor,colortbl} 
\graphicspath{ {./images/} } 
\usepackage{array} 
\usepackage{enumitem}
\usepackage{tabularx}
\usepackage{etoolbox}
\newcolumntype{P}[1]{>{\centering\arraybackslash}p{#1}}
\AtBeginEnvironment{table}{%
    \setlist[itemize]{nosep,     
                      leftmargin = *         ,
                      label      = $\bullet$ ,
                      before     = \vspace{-0.6\baselineskip},
                      after      = \vspace{-\baselineskip}
                        }
                           }
                           
\usepackage{authblk}

\title{ Evaluation of software impact designed for biomedical research: Are we measuring what's meaningful?}

\begin{document}

\author[1,2]{\bf Awan Afiaz} 
\author[3]{\bf Andrey A. Ivanov} 
\author[4]{\bf John Chamberlin} 
\author[5]{\bf David Hanauer} 
\author[2]{\bf Candace L. Savonen} 
\author[6]{\bf Mary J. Goldman} 
\author[7]{\bf Martin Morgan} 
\author[8]{\bf Michael Reich} 
\author[9]{\bf Alexander Getka} 
\author[10,11,12,13]{\bf Aaron Holmes} 
\author[9]{\bf Sarthak Pati} 
\author[10,11,12,13]{\bf Dan Knight} 
\author[10,11,12,13]{\bf Paul C. Boutros} 
\author[9]{\bf Spyridon Bakas} 
\author[14]{\bf J. Gregory Caporaso} 
\author[15]{\bf Guilherme Del Fiol} 
\author[16]{\bf Harry Hochheiser} 
\author[17]{\bf Brian Haas} 
\author[18]{\bf Patrick D. Schloss} 
\author[19]{\bf James A. Eddy} 
\author[19]{\bf Jake Albrecht} 
\author[20]{\bf Andrey Fedorov} 
\author[21]{\bf Levi Waldron} 
\author[2]{\bf Ava M. Hoffman} 
\author[15]{\bf Richard L. Bradshaw} 
\author[2]{\bf Jeffrey T. Leek} 
\author[2,*]{\bf Carrie Wright} 
\affil[1]{Department of Biostatistics, University of Washington,Seattle, WA}
\affil[2]{Biostatistics Program, Public Health Sciences Division, Fred Hutchinson Cancer Center,Seattle, WA}
\affil[3]{Department of Pharmacology and Chemical Biology, Emory University School of Medicine, Emory University, Atlanta, GA}
\affil[4]{Department of Biomedical Informatics, University of Utah, Salt Lake City, UT}
\affil[5]{Department of Learning Health Sciences, University of Michigan Medical School, Ann Arbor, MI}
\affil[6]{University of California Santa Cruz, Santa Cruz, CA}
\affil[7]{Roswell Park Comprehensive Cancer Center, Buffalo, NY}
\affil[8]{University of California, San Diego, La Jolla, CA}
\affil[9]{University of Pennsylvania, Philadelphia, PA}
\affil[10]{Jonsson Comprehensive Cancer Center, University of California, Los Angeles, CA}
\affil[11]{Institute for Precision Health, University of California, Los Angeles, CA}
\affil[12]{Department of Human Genetics, University of California, Los Angeles, CA}
\affil[13]{Department of Urology, University of California, Los Angeles, CA}
\affil[14]{Pathogen and Microbiome Institute, Northern Arizona University, Flagstaff, AZ}
\affil[15]{Department of Biomedical Informatics, University of Utah, Salt Lake City, UT}
\affil[16]{Department of Biomedical Informatics, University of Pittsburgh,Pittsburgh, PA}
\affil[17]{Methods Development Laboratory, Broad Institute, Cambridge, MA}
\affil[18]{Department of Microbiology and Immunology, University of Michigan, Ann Arbor, MI}
\affil[19]{Sage Bionetworks, Seattle, WA}
\affil[20]{Department of Radiology, Brigham and Women's Hospital, Harvard Medical School, Boston, MA}
\affil[21]{Department of Epidemiology and Biostatistics, City University of New York Graduate School of Public Health and Health Policy, New York, NY}
\affil[*]{Correspondence to \nolinkurl{cwright2@fredhutch.org}}

\maketitle

\begin{abstract}
Software is vital for the advancement of biology and medicine. Through analysis of usage and impact metrics of software, developers can help determine user and community engagement. These metrics can be used to justify additional funding, encourage additional use, and identify unanticipated use cases. Such analyses can help define improvement areas and assist with managing project resources. However, there are challenges associated with assessing usage and impact, many of which vary widely depending on the type of software being evaluated. These challenges involve issues of distorted, exaggerated, understated, or misleading metrics, as well as ethical and security concerns.  More attention to the nuances, challenges, and considerations involved in capturing impact across the diverse spectrum of biological software is needed. Furthermore, some tools may be especially beneficial to a small audience, yet may not have comparatively compelling metrics of high usage. Although some principles are generally applicable, there is not a single perfect metric or approach to effectively evaluate a software tool’s impact, as this depends on aspects unique to each tool, how it is used, and how one wishes to evaluate engagement. We propose more broadly applicable guidelines (such as infrastructure that supports the usage of software and the collection of metrics about usage), as well as strategies for various types of software and resources. We also highlight outstanding issues in the field regarding how communities measure or evaluate software impact. To gain a deeper understanding of the issues hindering software evaluations, as well as to determine what appears to be helpful, we performed a survey of participants involved with scientific software projects for the Informatics Technology for Cancer Research (ITCR) program funded by the National Cancer Institute (NCI). We also investigated software among this scientific community and others to assess how often infrastructure that supports such evaluations is implemented and how this impacts rates of papers describing usage of the software. We find that although developers recognize the utility of analyzing data related to the impact or usage of their software, they struggle to find the time or funding to support such analyses. We also find that infrastructure such as social media presence, more in-depth documentation, the presence of software health metrics, and clear information on how to contact developers seem to be associated with increased usage rates.  Our findings can help scientific software developers make the most out of the evaluations of their software so that they can more fully benefit from such assessments.
\end{abstract}


\section{Introduction} Biomedical software has become a critical component of nearly all biomedical research.  The field has generally embraced open science practices, including sharing and harmonizing data and creating open source (publicly accessible and often freely available) software \cite{green_strategic_2020, levet_developing_2021, itcr_open-source_2021, merow_better_2023}. Often such open source software is initially developed so that the developers can use it themselves to reach a research goal \cite{bitzer_intrinsic_2007}. The software is then shared with the hopes that the use of the software by others will also contribute to the advancement of science or healthcare. Once shared, evaluation of the software is necessary to achieve two major benefits: 1) inform developers about how to improve the use and ultimately impact of the tool and 2) demonstrate the value of the tool to support the developers to obtain funding to continue supporting existing software and creating new software. Common metrics include number of new users, returning users, and total downloads of the software, but the types of possible metrics vary based on the type of tool and the context. See Table ~\ref{tab:metrics_table}. These and other metrics can allow assessment of the rate of uptake and establishment within a community. However, developers often lack knowledge about the infrastructure or tools that can aid in the effective collection of usage and impact metrics. Proper metrics should not only examine the software's performance but assess whether motivations and goals of the researchers who are using the software are being met. Furthermore, metrics should also attempt to gather information about the downstream impact of the tool. 

\begin{table}[!ht]
 \caption{Common Metrics}
  \centering
  \begin{tabular}{|p{0.2\textwidth}|p{0.25\textwidth}|p{0.28\textwidth}|}
    \hline
    \multicolumn{1}{|c|} {\cellcolor[gray]{.9} \textbf{Measure}} 
    & \multicolumn{1}{|c|} {\cellcolor[gray]{.9} \textbf{Example Metrics}}     
    & \multicolumn{1}{|c|} {\cellcolor[gray]{.9} \textbf{Use}}  \\
    \hline
    \multirow{4}{*}{Tool Dissemination} 
    & 
     \begin{itemize}
         \item Total unique downloads
         \item New users
         \item Returning users
         \item Download count by version
     \end{itemize} & 
    \begin{itemize}
         \item Determining popularity of a given tool
    \end{itemize}  \\ 
           \cline{2-3}
    &     
    \begin{itemize}
         \item Download count by version
    \end{itemize}  &
    \begin{itemize}
         \item Assessing if users are keeping up-to-date
 \end{itemize}  \\ 

    \hline
    Tool Usefulness & \raggedright{
    \begin{itemize}
         \item Number of software engagements by user
    \end{itemize} 
    } &
    \begin{itemize}
         \item Determining prevalence of usage
    \end{itemize} \\
    \hline
    Tool Reliability & 
    \begin{itemize}
         \item Proportion of runs without a crash or error
    \end{itemize}  &
    \begin{itemize}
        \item Improving error handling, bug fixes 
    \end{itemize} \\
    \hline
    Tool Versatility &
        \begin{itemize}
        \item Distribution of data types (inferred from metadata) 
        \end{itemize} &
        \begin{itemize}
        \item Improving tool flexibility \& generalizability
        \end{itemize}\\
    \hline
    Interface Acceptability &
    \begin{itemize}
        \item Proportion of visitors who actually engage with the tool
        \item User error frequency
    \end{itemize} &
    \begin{itemize}
        \item Graphical tool and website acceptability
        \end{itemize}\\
    \hline
    Performance &
    \begin{itemize}
        \item Maximum memory usage
        \item Average time-to-complete of algorithmic steps 
    \end{itemize} &
    \begin{itemize}
        \item{Requirements analysis}
        \item{Tuning}
    \end{itemize}\\
    \hline
  \end{tabular}
  \label{tab:metrics_table}
\end{table}

We aim to provide guidance for evaluations of software impact and engagement so that software developers can build tools that better meet researchers' needs to catalyze the progress of biomedical research. We also discuss ethical considerations and challenges of such evaluations that still require solutions. The guidance presented here holds the potential for developers to improve the use and utility of their tools, improve their chances of funding for future development, and ultimately lead to the development of even more useful software to advance science and medicine \cite{wratten_reproducible_2021}.

\section{Assessment of attitudes and practices for software impact evaluations}

We performed two analyses to get a better sense of the landscape of software evaluation within the community of developers of the Informatics Technology for Cancer Research (ITCR) program funded by the National Cancer Institute (NCI). Our first analysis aimed to better understand how developers think about software evaluation by performing a survey. Our second analysis aimed to determine what infrastructure is often implemented to support software evaluation and if such implementation was associated with the frequency of papers describing usage of the software. See Supplemental Note \ref{sec-supp-note-methods} for more information about the methods. 

To perform the first analysis we surveyed 48 participants of the ITCR community. The survey asked questions about attitudes and practices of those involved with software development, maintenance, or outreach (in which respondents could often select more than one response). It was determined that limited time (68\% of respondents) and funding (57\% respondents) were the major barriers for performing software impact evaluations. Indeed although there are a few funding mechanisms that support the maintenance and analysis of software (as opposed to creation of new software), such as the Informatics Technology for Cancer Research (ITCR) program funded by the National Cancer Institute (NCI) sustainment awards \cite{kibbe_cancer_2017, warner_informatics_2020}, or the Essential Open Source Software for Science program of the Chan Zuckerberg Initiative \cite{CZ_essential_2019}, there is much more need for such funding for software sustainability compared to what is currently available. Indeed awareness of such a need was recently demonstrated by the recent Declaration on Funding Research Software Sustainability by the Research Software Alliance (ReSA) \cite{barker_amsterdam_2023}. While scientific software has become critical to most researchers, the funding to support the maintanence of such scientific software is not reflective of the current level of usage \cite{siepel_challenges_2019}. The next major barriers were privacy concerns (38\% of respondents), technical issues (32\% of respondents), and not knowing what methods to use for such evaluations (27\% of respondents). Despite these apparent challenges,  73\% of respondents state that such evaluations have informed new development ideas, 60\% stated that it informed documentation, and 54\% stated that it helped justify funding.  Thus additional support for evaluations of software usage and impact could greatly benefit the continued development of software. Responses to an open-ended question asking "Is there anything you would like to measure but have been unable to capture?" included (each of these examples were unique responses): collaborations that the tool supported, the number of commercial applications using the tool, the fraction of assumed user base that actually uses the tool, the downstream activity - what do users do with the results, and user frustration. These responses outline many of the challenges that developers often face in their tool evaluations. See Supplemental Table ~\ref{tab:survey_table} for examples of the goals the respondents had in analyzing their software.

\begin{figure}[ht] 
    \centering
\includegraphics[width=\textwidth,height=\textheight,keepaspectratio]{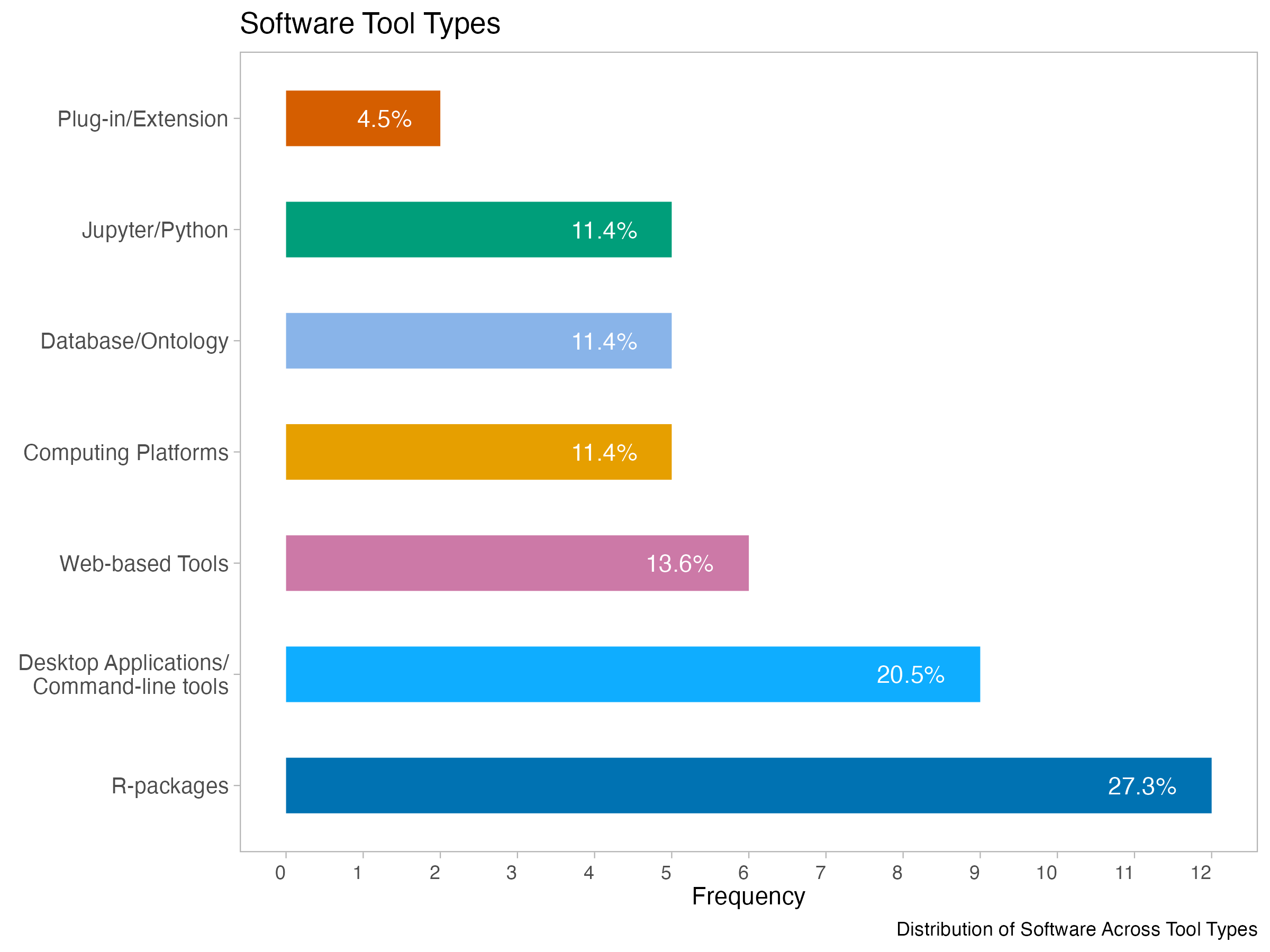}
    \caption{Variety of the 44 ITCR and CTD² tools evaluated for various characteristics with manual inspection}
    \label{fig:tool_eval}
\end{figure}

We also manually inspected 44 software tools, 33 of which were funded by ITCR alone and 7 funded by the Cancer Target Discovery and Development (CTD²) Network \cite{aksoy_ctd2_2017}, as well as 4 tools funded by both.  Each were inspected for aspects related to infrastructure that could help users know about the software tool or how to use it, as well as infrastructure related to software health metrics, to indicate to users how recently the code was rebuilt or tested, and investigated if there were any associations with these aspects and usage.  A variety of different types of research-related software tools were inspected (from R packages to platforms that allow users access to computing and data resources) - See Figure \ref{fig:tool_eval}. Each tool was manually inspected (by someone not involved in developing these tools) to get the experience of a potential user briefly examining related websites to determine if the tool had: a DOI for the software itself, information on how to cite the software, information on how to contact the developers, documentation (and how much), a twitter presence, and badges about health metrics visible on a related website.

\begin{figure}[ht]
    \centering
\includegraphics[width=\textwidth,height=\textheight,keepaspectratio]{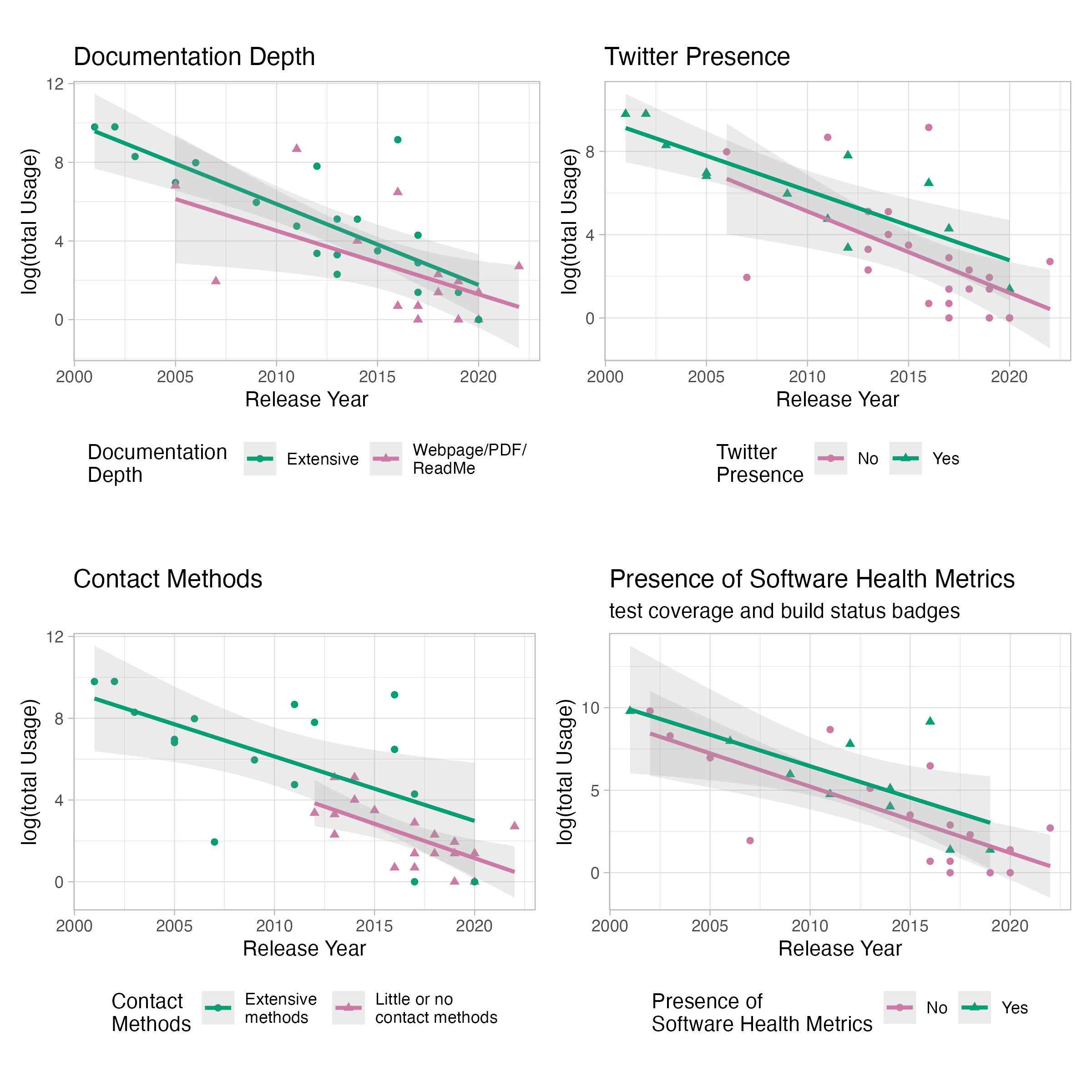}
    \caption{Aspects of software infrastructure appear to be associated with a larger number of published manuscripts from users describing usage of the software in the SoftwareKG-PMC database. The X-axis indicates the age of the software by showing the year that it was released. They Y-axis indicates the log of the total number of papers that describe usage of the software in the SoftwareKG-PMC database. See the supplement and our \href{https://hutchdatascience.org/ITCR_Metrics_manuscript_website/}{website} for more information.}
    \label{fig:inf_cit}
\end{figure}

To evaluate usage, we used the SoftwareKG-PMC database \cite{Kruger_usage_20}, which does not include citations to tools, only plain-text mentions inferred by a text-mining algorithm. The database does not know anything about these tools per se, and not all of these mentions necessarily correspond to the same tool. For example, DANA is an ITCR tool for microRNA analysis but there are also other unrelated methods with the same name. On the other hand, tools with highly specific names like Bioconductor are unlikely to have the same issue. Most notable was a finding that although time since the tool release is the largest contributor to variation in the number of papers describing usage, various aspects of infrastructure that could help users know about a tool (social media on twitter), have confidence in the tool (badges about software builds or tests on code repositories), or learn more about how to use the tool (extensive documentation and feedback mechanisms) all seem associated with an increased rate of manuscripts that describe using the tool. All show significant association (p<0.05) with usage when not accounting for tool age. Only having extensive feedback mechanisms was significantly associated when also accounting for tool age. See Figure \ref{fig:inf_cit}. For more information about this analysis, see this \href{https://hutchdatascience.org/ITCR_Metrics_manuscript_website/}{website}.

\section{Overall guidance}

\subsection{Successful evaluations are anchored by an understanding of the intended use of the software}
\label{sec:use_understanding}
The intended goal or purpose of the scientific software should be used to inform how the software is evaluated \cite{Basili94}. Computational tools are designed to support well-defined goals often called use cases \cite{gamma_design_1995} for specific sets of users called personas\cite{cooper_inmates_2004}. Efforts to evaluate the impact of tools should be guided by a clear understanding of these use cases and personas to assess how well the tools meet the intended goals. 

As an example, when the intended purpose of the software is to contribute to the treating, diagnosing, or prevention of a medical condition, it may qualify as a "Medical Device" that requires clinical validation. Clinical validation includes:

\begin{quote}
\textit{... a systematic and planned process to continuously generate, collect, analyze, and assess the clinical data pertaining to a [the software] in order to generate clinical evidence verifying the clinical association and the performance metrics of a [the software] when used as intended by the manufacturer. The quality and breadth is determined by the role of [the software] for the target clinical condition, and assures that the output of [the software] is clinically valid and can be used reliably and predictably.}\cite{clinical_evaluation_2017}
\end{quote}

\subsection{Metric selection should be hypothesis driven} 
\label{sec:hypothesis_driven}
Collecting an exhaustive amount of user data, and then selecting metrics, can add complexity and increase the risk that metrics are selected in a biased manner. To mitigate this, hypothesis-driven metrics can be selected ahead of time based on a specific hypothesis to ultimately evaluate how well the software supports its intended goals \cite{design_driven_dev_2020}.

\subsection{Intentions for evaluation can also inform design choice} Software can also be designed with future evaluations in mind. Once the intended use of the software is clearly defined, user interfaces can be iterated to effectively collect the right data. For example, Xena \cite{xena_2020} is a tool intended to enable users to visualize various aspects of the genome. The developers collect metrics involving how often users use the tool to perform visualizations. However, consideration of privacy led the developers to not collect metrics about what part of the genome gets visualized.

\subsection{Metrics can achieve different goals for different audiences}

Clear understanding of which use cases, personas, and audience(s) are of interest, as well as what motivations are of interest, can help guide what user metrics to collect to achieve project goals. For example, if the audience is the developers themselves, and the motivation is to learn how to retain new users, it may be helpful to understand how well new users are able to learn the basics of the tool. Thus focusing on metrics related to interactions with tutorials may be the most useful. It is also worthwhile to consider which presentations of software metrics will be most compelling to the intended audience. Detailed metrics on the inner workings of a software tool might be highly-informative to developers, but such metrics would be far too granular to be of interest to funding agencies.

Developers might be interested in tool optimization, which can involve improving workflows, performance, usage, or usability. For example, recognizing the types or volumes of data being used, as well as temporal trends of data usage, can highlight opportunities for new algorithm development. Such evaluations can guide future work \cite{fenton2014software}.

External audiences might instead be more interested in evidence for impact. This is often required for developers to gain resources to develop or maintain semi- or un-related tools. Demonstration of past capability to develop impactful software supports requests to do so in the future. Demonstrating that a tool is widely-used and widely-accepted can also encourage users to adopt a tool more readily, and be more invested in a tool community. Developers may be drawn to projects with impact to build upon an exiting tool. Finally, demonstration of impact can recruit new users which can diversify tool communities and bring new problems of interest that expand the utility of a tool. See Table \ref{tab:benefit_table} for more details.
 
\begin{table}[!ht]
 \caption{\textbf{Software evaluation supports internal needs for tool optimization and development, as well external needs to demonstrate tool value 
 to others.} }
  \centering
  \begin{tabular} {|p{0.27\textwidth}|p{0.26\textwidth}|p{0.39\textwidth}|}
    \hline
    \multicolumn{1}{|c|}{\cellcolor[gray]{.9} \textbf{Internal Need}} 
    & \multicolumn{1}{|c|}{\cellcolor[gray]{.9} \textbf{Specific Goal}}
    & \multicolumn{1}{|c|}{\cellcolor[gray]{.9} \textbf{Benefit}}\\[1.1ex]
    \hline
    \multirow{17}{*}{Tool Optimization}               
    & \multirow{3}{*}{Improving Workflows} & 
    Identify unexpected usage \\
    & &
    Identify code inefficiencies \\
    & &
    Identify resource usage inefficiencies \\
    &&
    Identify inadequate documentation\\ \cline{2-3}
    &   \multirow{4}{*}{  }
    & 
     Identify mismatches with defaults and use \\
     &  Improve Performance  &
    Assess user wait times \\
    &  & 
     Measure data volume \\  \cline{2-3}
    & \multirow{5}{*}{ Improve Usage} & 
    Identify software errors \\
    &&
    Identify what features are used and not used \\
    &&
    Identify who the user-base is \\
    & &
    Determine user-base diversity \\
    & &
    Identify sources of other possible users \\
    & &
    Determine what users expectations are \\
    & &
    Determine if user expectations are appropriate \\
    & &
    Evaluate success of outreach approaches\\  \cline{2-3}
    & \multirow{2}{*}{ Improve Implementation} & 
    Identify barriers for adoption \\
    &   &
    Identify methods to support adaption \\
    &  & 
    Identify use of out-dated versions\\\cline{2-3}
    & \multirow{2}{*}{ Improve Usability} & 
    Identify user errors \\
    & & 
    Identify if and how users are struggling \\ 
    \hline
    \multirow{6}{*}{\shortstack{Tool Development \\ \& Maintenance}}
    & \multirow{6}{*}{\shortstack{ Guide Future Work\\  Motivate Continued Support}} &
    Enumerate data types being used \\
    & &
    Discover opportunities for new features \\
    && 
    Discover data needed to address user goals  \\
    & & 
    Identify more appropriate resource allocation \\

    \hline
    \multicolumn{1}{|c|}{\cellcolor[gray]{.9} \textbf{External Need}} 
    & \multicolumn{1}{|c|}{\cellcolor[gray]{.9} \textbf{Specific Goal}}
    & \multicolumn{1}{|c|}{\cellcolor[gray]{.9} \textbf{Benefit}}\\[1.1ex] 
    \hline
    \multirow{4}{*}{Gain Support}              
    & \multirow{4}{*}{Show Evidence of Impact} & 
    Support future funding requests \\
    & &
    (to maintain or develop new tools) \\
    & &
    Request for resources  \\
    & & 
    (to maintain or develop new tools) \\[1.1ex]
    \hline
    \multirow{4}{*}{Gain User Commitment} 
    & \multirow{4}{*}{Evidence of Tool Acceptance} & 
    Reassure users about tool to:\\
    &&
    - Promote continued use\\
    && 
    - Promote usage of new tools by the same developers \\ 
    &&
    - Promote usage by more diverse users \\[1.1ex]
    \hline
    \multirow{1}{*}{Gain Community Development} 
    & \multirow{1}{*}{Evidence of Co-Development} & 
    Encourage contributions  \\
    \hline
  \end{tabular}
  \label{tab:benefit_table}
\end{table}

\subsection{No single evaluation method works for every type of software}
\label{sec:no_one_way}
No individual scheme for collecting metrics fits every type of research software tool.  The meaning of a set of metrics may differ across contexts. The location of a tool (e.g., whether it is on the web or downloaded) can affect metric collection and influence user access to software versions. For example, for a web-based application, it may be feasible to collect user metrics on a per-account basis and to collect information about the type of data or software features users tend to work with. With web-based tools, users will rarely have access to older versions of the software. Thus developers can add version updates and collect metrics with clarity about how usage changed with updates. On the other hand, for tools that must be run (and possibly installed) locally, users may be using older versions of the software that they previously downloaded. Here, metrics on a per-version basis provide a much better representation of usage, rather than simply the overall number of unique downloads. Collection of version usage is important, as developers may want to pair this with citation data to know if users are relying on out-of-date aspects of previous versions of the software.

 \subsection{Metrics should be interpreted}
 
 Interpretation of user metrics can be tricky since any given metric may have many obvious or subtle influences. When software has sufficient use, observed spikes in usage, both up and down, provide important feedback. A spike may correspond to a class or workshop using the tool or a recent publication that cites the tool. Negative trends in usage may indicate a break in the academic calendar, down time of a host server, software bugs or, for tools based on Amazon Web Services, the additional compute resources required by Amazon during holiday cycles may preempt software running on spot instances. It is also important to avoid comparisons between metrics for tools with different users. For example, clinical tools that require institutional approvals and support will have lower installation rates than other software tools. Tools that are very useful to a small number of users may still have important impact on the field. In such cases qualitative metrics may be useful or different quantitative metrics about usage among the smaller user base.  
 
The total unique downloads might be useful as a metric of software popularity, but it only reflects how many people have tried to download it and not if users find it useful. Instead, one might consider measuring how often users interact with software or counting the number of launches of the software which run over a certain predefined session time threshold to better evaluate actual usage. Returned usage by the same users often suggests that the users may find the tool useful. For tools that offer access or analyses of different data types, one may want to parse usage by data types to evaluate how useful the tool appears to support different kinds of users.   Specific measures can provide a common basis comparing versions and potentially against other similar software.

\subsection{Software infrastructure enables impact evaluation}
There are several components of a software tool that can assist with the assessments of the software impact and engagement. Once a developer team better understands their audience, use cases, personas, and assessment motivations, the infrastructure described in Table~\ref{tab:inf_table} and Supplemental Note ~\ref{sec-supp-note-inf} could benefit or enable such evaluations, as well as improve user awareness and engagement.

\begin{table}[!ht]
 \caption{\textbf{Software infrastructure enables the capture of valuable metrics for evaluating engagement and impact.} Note that there are other helpful tools to enable metric collection. These are simply examples.}
  \centering
  \begin{tabular} {|p{0.1\textwidth}|p{0.16\textwidth}|p{0.17\textwidth}|p{0.19\textwidth}|p{0.24\textwidth}|}
    \hline
    \multicolumn{1}{|c|}{\cellcolor[gray]{.9} \textbf{Elements}} 
    & \multicolumn{1}{|c|}{\cellcolor[gray]{.9} \textbf{Options}}
    & \multicolumn{1}{|c|}{\cellcolor[gray]{.9} \textbf{\shortstack{\\Tools to Enable \\Metric Collection}}}
    & \multicolumn{1}{|c|}{\cellcolor[gray]{.9} \textbf{\shortstack{\\Possible\\ Enabled Metrics}}}
    & \multicolumn{1}{|c|}{\cellcolor[gray]{.9} \textbf{Considerations}}\\[1.1ex]
    \hline
    \multirow{8}{*}{\shortstack{Web \\ Presence}}               
    &   \multirow{2}{*}[-2em]{ Web-based tool} & 
    \raggedright{
    \begin{itemize}
        \item Cronitor \cite{cronitor} for tools using cron job scheduling \cite{cron_2009})
        \item Google Analytics \cite{google_analytics}
    \end{itemize}
    }
    &  \raggedright{ 
    \begin{itemize}
    \item Identify details about usage
    \item Identify where your tool is being used
    \item Possibly identify what data are being used
    \end{itemize}
    } &  May need to consider privacy restrictions for tracking IP addresses. \\
    \cline{2-5}
    & \multirow{2}{*}[-1em]{ \shortstack{Documentation \\ Website}} &
    \begin{itemize}
        \item Google Analytics \cite{google_analytics}
    \end{itemize}
    & \raggedright{ 
    \begin{itemize}
    \item Counts of page views and scrolls
    \end{itemize}
    } &
    Pages with more views may identify widely used features or confusing aspects. \\
    \hline
    \multirow{3}{*}[-2em]{Citability}
    & \raggedright{Providing something to cite (Software DOI or manuscript) and information on how to cite} & \raggedright{
    \begin{itemize}
        \item To create DOIs: Zenodo \cite{zenodo}, Dryad \cite{datadryad}, Synapse \cite{synapse}, and Figshare \cite{figshare} 
        \item To track DOIs: Altmetric \cite{noauthor_altmetric_2015}
    \end{itemize}} & \raggedright{ 
     \begin{itemize}
         \item Total citation counts 
         \item Counts of citations by journals of different fields
     \end{itemize}
     } & Semantic Scholar \cite{noauthor_semantic_nodate} provides reports that indicate where citations have occurred within scientific articles.\\
    \hline
    \multirow{3}{*}[-5em]{Contact} &
    \multirow{3}{*}{\shortstack{\\Feedback \\Mechanisms}} & 
    \raggedright{
    \begin{itemize}
    \item GitHub Issue Templates
    \item Surveys
    \end{itemize}
    }
    & \raggedright{
    \begin{itemize}
    \item User feedback count
    \item Addressed user feedback count 
    \end{itemize}
    } & Often users will only provide feedback if something is broken. Depending on the tool, many users may not be comfortable with GitHub Issues.\\
    \cline{2-5}
    & \multirow{2}{*}{ \shortstack{Discussion \\ Forums}} & \raggedright{ 
    \begin{itemize}
        \item Discourse \cite{discourse}
        \item Biostar \cite{biostars}
        \item Bioinformatics Stack Exchange \cite{bioinformaticsstackexchange}
        \item Google Groups \cite{google_groups}
    \end{itemize}
    } & \raggedright{
    \begin{itemize}
        \item Metrics based on user engagements and answered questions 
    \end{itemize}
    } &
    Forums save time for development as users help each other instead of developers answering individual emails for repeat problems. A code of conduct can help create a supportive community.\\
    \cline{2-5}
    & \multirow{2}{*}{ \shortstack{Newsletter\\Emails}} &
    \begin{itemize}
        \item Mailchimp \cite{mailchimp}
        \item HubSpot \cite{hubspot} 
    \end{itemize} & \raggedright{ 
    \begin{itemize} 
        \item Count of newsletter openings
        \item Count of link clicks
        \item Count of unsubscribers
    \end{itemize}
    } & Newsletters can help inform users about new features. \\
    \hline
    \multirow{3}{*}{\shortstack{Usability\\Testing}}
    & \raggedright{Observe a few people use the tool} & 
    \raggedright{
    \begin{itemize}
        \item Zoom screen sharing and recording
    \end{itemize}
    } & \raggedright{
    \begin{itemize} 
        \item Qualitative information about how users interact with your software
    \end{itemize}
    } & Even low numbers of usability interviews (~3) can yield fruitful lessons that can be paired with other metrics to guide development.\\
    \hline
    \multirow{3}{*}{Workshops}
    & \raggedright{ 
    \begin{itemize}
        \item Online or in-person
        \item Basics or new features
    \end{itemize}
    } & \raggedright{ 
    \begin{itemize}
        \item Attendees can participate in surveys
    \end{itemize}
    } & \raggedright{ 
    \begin{itemize} 
        \item Quantity, duration, and attendance at workshops are metrics that can be reported to funding agencies
    \end{itemize}
    } & Recordings can be posted on Social Media (for additional metrics).\\
    \hline
    \multirow{3}{*}{\shortstack{Social\\ Media}}
    & \raggedright{ 
    \begin{itemize}
        \item YouTube Videos
        \item Twitter/ Mastodon
        \item Instagram
        \item LinkedIn 
    \end{itemize}
    } & \raggedright{
    \begin{itemize}
        \item Hootsuite \cite{hootsuite} - social media management
    \end{itemize}
    } & \raggedright{ 
    \begin{itemize} 
        \item Engagement metrics (video watch counts, likes, etc) 
    \end{itemize}
    } & Pairing Social media metrics with software engagement metrics can determine if outreach strategies are successful. \\
    \hline
       \multirow{2}{*}{Reviews}
    & \multirow{2}{*}{Review Forum} & 
    \begin{itemize}
        \item SourceForge
        \item GitHub
    \end{itemize} &  \begin{itemize}
        \item Stars
        \item Number of reviews
    \end{itemize} & Positive reviews can reassure new users and funders.\\ 
    \hline
  \end{tabular}
  \label{tab:inf_table}
\end{table}

\subsection{Software project health metrics can reassure users and funders}
Tracking adherence to standards of software engineering can be a useful way to assess software project health including the use of version control systems, high coverage of code with testing, and use of automated or continuous integration. None of these measures of project health are perfect (and they can be done poorly) but overall they can be collectively assessed as indicators of software health. Including badges for such indicators on code repositories and websites can give users and others confidence in your tools.  Additional detail on these topics, can be found in The Pragmatic Programmer\cite{thomas_pragmatic_2019}. See Table~\ref{tab:soft_health_table} and Supplemental Note ~\ref{sec-supp-note-health-inf} for suggestions.

\begin{table}[!ht]
 \caption{\textbf{Software health infrastructure enables collecting metrics that can reassure users and funders.}  }
  \centering
  \begin{tabular} {|p{0.1\textwidth}|p{0.135\textwidth}|p{0.12\textwidth}|p{0.21\textwidth}|p{0.22\textwidth}|}
    \hline
    \multicolumn{1}{|c|}{\cellcolor[gray]{.9} \textbf{Infrastructure}} 
    & \multicolumn{1}{|c|}{\cellcolor[gray]{.9} \textbf{Options}}
    & \multicolumn{1}{|c|}{\cellcolor[gray]{.9} \textbf{\shortstack{\\Tools to Enable \\Metric Collection}}}
    & \multicolumn{1}{|c|}{\cellcolor[gray]{.9} \textbf{\shortstack{\\Possible\\ Enabled Metrics}}}
    & \multicolumn{1}{|c|}{\cellcolor[gray]{.9} \textbf{Considerations}}\\[1.1ex]
    \hline
    \multirow{6}{*}{\shortstack{Version\\ Control}}              
    &   \multirow{3}{*}[-3em]{\shortstack{Without\\ Automation}} & \raggedright{\begin{itemize}
        \item Git/GitHub \cite{GitHub} (The insight tab and API allow for systematic metric collection)
        \item Git/GitLab \cite{GitLab}
        \item BitBucket \cite{bitbucket}
    \end{itemize}}
    & \raggedright{
    \begin{itemize}
    \item Commit frequency (how often code is updated)
    \item Date of the most recent commit
    \item Number of active contributors
    \item Software versions updates
    \end{itemize}
    } &
     Commit frequency allows assessment of how actively the software is being maintained. The number of contributors can indicate sustainability. One single contributor may pose a sustainability risk. Version information can enable users to determine if they are using the most up-to-date version. \\
    \cline{2-5}
    & \multirow{5}{*}[-3em]{ \shortstack{With\\ Automations}} & \raggedright{
    \begin{itemize}
        \item GitHub Actions\cite{github_actions}
        \item Travis CI \cite{Travis}
        \item CircleCI \cite{circleCI}
    \end{itemize}
    }
    &
    \begin{itemize}
    \item  Current build status (if the software built without errors) 
    \end{itemize} &
    Continuous Integration and Continuous Deployment or Delivery are terms to describe 
 a situation where every time code is modified, the full code product is automatically rebuilt or compiled. Continuous Deployment or Delivery describes the automatic release of this new code to users. Delivery in this case describes situations where the software requires more manual releases while deployment is seamless. GitHub Actions can also help with metric collection from the GitHub API. \\
    \hline
    \multirow{3}{*}[-2em]{Testing} 
    & \multirow{2}{*}[-3em]{ \shortstack{Automated \\ Testing}} & \raggedright{
    \begin{itemize}
        \item GitHub Actions \cite{github_actions}
    \end{itemize}
    } & \raggedright{
    \begin{itemize} 
    \item Test code coverage (the fraction of lines of code in the project that are covered by tests)
    \end{itemize}
    } & Unit tests check individual pieces of code; component and integration tests check that pieces of code behave correctly together; acceptance tests check the overall software behavior. Achieving in-depth test coverage requires careful software design. Test coverage does not evaluate the quality of the test cases or assertions. \\
    \hline
    \multirow{3}{*}[-2em]{Licensing}
    & \raggedright{A variety of licenses exist to allow or disallow reuse and to require attribution} & \raggedright{
    \begin{itemize}
        \item Creative Commons \cite{creative_commons}
    \end{itemize}
    } &
    \begin{itemize}
        \item Possible quantification of reuse of your software code
    \end{itemize} & Clearly indicating if and how people can reuse your code will make them more comfortable to do so. Determining when this is done can be a challenge, but requiring attribution makes this more feasible\\
    \hline
  \end{tabular}
  \label{tab:soft_health_table}
\end{table}

\subsection{Metrics related to software quality and re-usibility could reassure users and funders}

 Software re-usability metrics have been suggested to enable better discernment of the capacity for code to be reused in other contexts. These metrics can also evaluate if code is written to be more resilient over time to dependency changes and other maintenance challenges. One such example would be the degree to which aspects of the software are independent of one another \cite{mehboob_reusability_2021}. As research funders start to value software maintenance more, metrics related to resilience and re-usibility may become more valuable and could also encourage more resilient development practices.

\section{Challenges and nuances}

There are a number of challenges and nuances associated with evaluating metrics for software usage and impact. Here we outline some examples.

\subsection{Citation challenges}
Measuring the number of citations of your tool is especially useful as a metric to report to funding agencies. To enable this, it is necessary that your tool has a manuscript or other data object to cite \cite{chue_hong_software_2019}. Having a manuscript published as a preprint such as BioRxiv, or even in Figshare (see 'The graph-tool python library' \cite{peixoto_graph-tool_2017}), can help a tool to have a citable presence. For data or software objects, Zenodo \cite{zenodo}, Dryad \cite{datadryad}, Synapse \cite{synapse}, and Figshare \cite{figshare} can provide DOIs (Digital Object Identifiers), which allow citing the software more directly. 

Getting users to cite a tool or cite it correctly can be difficult. First, some tools are so common or fundamental that users often don't think to cite them, for example a tool like the UCSC Genome Browser \cite{ucsc, kent_human_2002}. Second, some tools are used in the discovery phase of a project, and a user may not think of it when they are in the final stages of writing up findings. An example would be a tool used to find a cohort of patients for further analyses, such as EMERSE (Electronic Medical Record Search Engine) \cite{hanauer_supporting_2015}. Third, tools which provide system architecture for other software may also not be typically cited. Some tools in this category include Bioconductor \cite{bioconductor}, Gene Pattern Notebook \cite{reich_genepattern_2017}, and Galaxy \cite{the_galaxy_community_galaxy_2022}. Understanding usage of these system level tools may require looking at usage of other tools that are available on these platforms.   

Lastly, while users may recognize and acknowledge your tool, they may not cite the tool in the reference section of their paper and they may mention the tool without complete information, such as not including version information, parameter settings, URLs, or credit to those who made the software. In fact, a study manually evaluating software citations of 4,971 academic biomedical and economics articles, found that the citations only included version information  28\% of the time \cite{howison_software_2016}.  Another study manually evaluating 90 biology articles, also finds a low rate of complete information with version information being included only 27\% of the time and URL information included only 17\% of the time \cite{du_softcite_2021}. People may mention a tool in a figure legend, in the paper itself, in the acknowledgments, or even in the abstract, without a citation. These mentions of a tool are difficult to track. Furthermore, occasionally manuscripts simply acknowledge that a software tool exists, rather than indicating that it was actually used by the authors. In other cases a newer version of a tool is used, yet a previous publication for an earlier version of the tool may often be cited by these users. This typically requires manually reading articles to discover the use of the software. 

Fortunately, there are a number of tools that can help measure citations. These include Google Scholar, Web of Science, PubMed, and ResearchGate. Additionally, some tools are being developed to help track software mentions, such as the tool "CORD-19 Software Mentions" \cite{wade_cord-19_2021}. Each of these citation measuring tools has benefits to overcome the above challenges. For instance, Google Scholar will allow you to search for the name of a tool anywhere in a paper and Semantic Scholar allows for reports of where citations are located across papers. Some tools names have more than one meaning depending on context, such as Galaxy, which can make it more difficult to use keywords to find citations. 

A couple of recent papers \cite{istrate_large_2022, schindler_role_2022} have developed automated extraction methods to overcome additional challenges, such as disambiguating multiple synonyms for the same software, typographic errors or misspellings, the use of acronyms vs full names, as well as capturing version information. It is also important to note that if other software relies on your software, it is likely useful to evaluate the citations of other such software in your analysis of the impact of your software. However, it can be difficult to know if this software exists if the developers do not adequately describe dependencies in a manuscript or documentation.

Software requires extensive work to maintain the utility (and sometimes security) over time. Typically it is much easier to publish manuscripts for a new piece of software. A lack of maintenance can be quite detrimental for research. Researchers do not want to waste time learning how to use software that no longer works. It takes valuable time to find a new solution for their analysis. A new system to reward updates with a new type of manuscript for software updates has been proposed \cite{merow_better_2023}. This could reduce issues of users not providing version information, reward developers who start working on software after the initial publication, and provide new ways for funding agencies and others to better recognize and reward software maintenance. 

\subsection{Limitations of tracking systems}

One other difficulty with the implementation of analytics platforms such as Google Analytics is that due to security and privacy concerns, some academic institutions are blocking connections to these services outright. Other, unknown or custom-built tracking may be flagged by security software or otherwise generically blocked. Hence, reliance on these data may not adequately capture industry or academic usage.

\subsection{Distorted metrics}

As metrics become more commonly used in broader applications than were originally intended, there is a risk that the meaning behind those metrics can become distorted. Projects like Bioconductor \cite{bioconductor}, with a large variety of software packages, offer an opportunity to assess the risk of this distortion by evaluating how packages are used over time. This can reveal important nuances about software usage metrics. See ~Table \ref{tab:dist_table}.

\begin{table}[!ht]
 \caption{Distorted Metrics}
  \centering
  \begin{tabular}{|P{0.2\textwidth}|p{0.65\textwidth}|}
    \hline
    \multicolumn{1}{|c|} {\cellcolor[gray]{.9} \textbf{Distortion}}     
    & \multicolumn{1}{|c|} {\cellcolor[gray]{.9} \textbf{Example}}  \\
    \hline
    \multirow{4}{*}{Accidental Usage} 
    & 
Occasionally scripts used on servers may inadvertently download a package repeatedly and rapidly hundreds to thousands of times, resulting in distorted download metrics that are not representative of real usage. Unique IP download information is useful to distinguish between one user downloading many times versus many users a few times. Given privacy concerns, an alternative solution could involve tracking the timing and approximate location of downloads with a threshold for what would be more than expected as maximum real usage, like a group of people following a tutorial.  \\
    
    \hline
\multirow{4}{*}{Background Usage} & 
  There is a baseline background level of downloads across all packages in Bioconductor (including those that are no longer supported). Thus if a new package has 250 downloads in the first year this may seem like a successful number, but actually it is similar to background levels. \\ 
    \hline
\multirow{4}{*}{\shortstack{Technical \\ vs Research usage}} & 
   The S4Vectors package \cite{S4Vectors}  is an infrastructure package used by many other packages for technical and non-biological reasons and is therefore not often directly downloaded by end-users. This package is also included in automated checks for other Bioconductor packages using GitHub actions. It can be difficult to discern if the usage of a package is for scientific research itself or supporting the implementation of other software. While both are arguably valuable, distinguishing between these motivations can help us understand a particular software's impact in a field.  \\ 
    \hline
\multirow{4}{*}{Usage Persistence} &
    The affy package \cite{affy}, was one of the early packages for microarray analysis, a technology that has largely been replaced by newer technologies, which can be seen by the rate of microarray submissions to GEO overtime. However, despite a the field transitioning away from microarray methods \cite{mantione_comparing_2014}, the package was downloaded in 2021 at rates that doubled the rates in 2011. The authors speculate that this could be due to people historically requesting that affy be installed on servers and that this is just persisting, or perhaps it is being used for preliminary hypothesis testing using existing micrarray data, or perhaps it is being used because other microarray packages are no longer supported.\\ 
    \hline
  \end{tabular}
  \label{tab:dist_table}
\end{table}

\subsection{Clinical data challenges}

Systems that contain clinical data have unique challenges.  Clinical data (generally defined as data extracted from electronic health records) often contain highly sensitive protected health information (PHI). While clinical details can be vital for research, the number of individuals that have access to the data is generally much smaller than tools designed to work with non-clinical data. It would not be realistic to compare usage metrics to more widely available and accessible tools. Many tools containing clinical data are also run at an enterprise level, meaning they are installed only one time by system administrators and accounts are provisioned to users. This affords greater control over the access to PHI and allows necessary auditing functions to record the viewing of patient information.  Thus, counting installations does not represent the overall use of the software. Unfortunately, it is not common for system administrators to send usage reports to software development teams interested in tracking usage.  Further, firewalls and built-in security mechanisms inhibit developers from accessing the installed systems themselves. Other approaches, such as software “phoning home” (the collection of information from the computers of users that downloaded or installed a particular software) could raise flags by security teams looking for unusual behavior, especially for malicious software that could be trying to send clinical data outside of the covered entity. The EMERSE \cite{hanauer_supporting_2015} GitHub repository, although open source, is now private for two reasons: (1) encourages  better understanding of who is interested in using EMERSE and development of relationships with such individuals and (2)  only known entities have access to the software to prevent those with malicious intentions from looking for vulnerabilities that may have been missed despite rigorous, continuous security reviews.  Ultimately, due to downloads typically being at an institutional level for clinical tools, metrics around software downloads would underestimate their potential impact.  

\subsection{Goodhart's law}
An important consideration for metrics for software assessments is Goodhart's Law, which states that “every measure which becomes a target becomes a bad measure\cite{hoskin_awful_1996}". As an example, h-indices (i.e., the number of papers an author has with that many or more citations) are often used to assess the quality of an author’s impact. However, as the h-index grew in popularity, the number of researchers included as co-authors, the number of citations per paper, and the fraction of self-citations increased. Each leading to an increased h-index. At the same time, these behaviors also increase a journal’s impact factor \cite{fire_over-optimization_2019}. Altmetric, described earlier, may help in providing information about more diverse engagements with articles (social media, news), however it does less to aid in evaluating author contribution. It is not a stretch to imagine that as metrics are developed and codified for tools, this would lead developers to attempt to improve the metric for their tool. Although Goodhart’s Law could be used to bring about best practices for binary outcomes (i.e., compliance with codes of conduct, public deposition of code), for more quantitative metrics (e.g., number of downloads, citations) the results would easily render the metrics meaningless. One way to avoid this dilemma is to continue to evaluate our metrics over time, consider if our metrics are truly measuring what we think they are, consider if our metrics are actually fair to a diverse range of project teams, and consider new metrics as needed \cite{fire_over-optimization_2019}. One example of such unfairness would be evaluations of metrics for clinical data resources that are inherently limited in terms of who can be allowed to access the resource. Such caveats need to be accounted for when evaluating and comparing tools and resources. Funding agencies need to consider how each type of tool is context-dependent, and that impact should be measured and compared between similar classes of tools with this in mind.

\subsection{Security, legal and ethical considerations}\label{sec:legal_ethics}
Often with the use of phone-home software or web-based analytics, users are tracked for not only downloads, but often for specific elements of usage, such as the number of times a user runs a particular kind of analysis, etc. Occasionally software developers will notify users that they are being tracked, however this is often not a requirement. The General Data Protection Regulation (GDPR), which became implemented in 2018, requires that organizations anywhere in the world respect certain data collection obligations regarding people in the European Union. It is intended to protect the data privacy of individuals and mostly guards against the collection of identifiable personal information. Thus, data collection of software usage needs to be mindful of the GDPR and any other international regulations that may impact data collection of users.  As science is particularly an international pursuit, often a majority of the users may reside outside the country where the tool was developed.

One option is to let users determine if they wish to be tracked by letting them know during certain stages of use depending on the type of software, such as when users download, register, or use certain features of the software. It is also possible for software developers to design tracking to be more anonymous. For example, a genome visualization tool may track the number of unique uses, but it will not track what part of the genome was visualized. Google Analytics also provides support for how to comply with such restrictions, for example you can  mask unique IP addresses of visitors to a website that tracked by the system \cite{google_analytics_privacy}.  Ethical and legal consequences should be considered when designing or implementing tracking systems of scientific software. See Supplemental Note \ref{sec-supp-note-sec} for more information.

\section{Discussion and conclusion:}

Overall our assessments indicate that cancer researchers of the ITCR find it difficult to find the time or funding to dedicate evaluating the impact and usage of their software, despite their awareness of the benefit of such evaluations. Many have found such evaluations useful for a number of reasons involving driving future development and with obtaining additional funding. We also find that a sizable portion (27\%) of researchers surveyed self-report as not knowing what methods to use for such evaluations. We hope that the guidance outlined here will be beneficial in informing software developers about such methods.  We also find that tools appear to be more widely used when software developers provide deeper documentation, badges about software health metrics on websites and repositories, and more in-depth contact information for users to reach the developers, as well as having a Twitter presence. It is not yet clear what is responsible for these findings. It may be that those who put more effort into their tools may create tools that are easier to use and therefore more widely used. However, it may also be that a Twitter presence brings new users to tools and that the other infrastructure (badges, deeper documentation, etc.) help new users to trust software, thus encouraging their usage. It would require more studies, perhaps from the user perspective, to further understand the patterns that we saw in our analyses. However, it does suggest that supporting software developers to spend more time on such elements could drive further usage of existing tools. We hope that funding agencies will value supporting developers to evaluate, promote, and maintain existing tools in addition to the current typical model for most agencies to prioritize the creation of new tools. A recent article \cite{merow_better_2023} suggested that a new type of manuscript for software updates may help the field to better reward development and maintenance of existing software. We argue that inclusion of evaluations of software impact and usage could also be incorporated into such a new model for software-related manuscripts. 

We also describe challenges and nuances associated with the evaluation of software usage and impact. We describe common distortions of metrics, ethical and security challenges in the collection of user data and the concept of Goodhart's law that metrics become a bad measure over time. We point out that citation rates may be lower for certain tools, such as clinical tools that require institutional support to implement or tools that are designed for using data or methods that a small number of researchers would use. Typical methods of assessments based on common metrics such as citations or number of users may underestimate the value of these tools to the field. While these metrics may be valuable for comparisons of similar types of tools, it is advisable that we also consider other types of metrics that may give more insight about the downstream impact of a tool. For example, perhaps we should consider how much a software tool inspires the development of other tools, the value of the papers that cite a tool (perhaps by citation rate, measures of innovation, or measures of clinical impact). Certainly as scientific software continues to be critical for scientific and medical advancement, this topic will only be of greater importance as we determine how to support scientific software developers in the future.

\section{Author Contributions}

Awan Afiaz and Carrie Wright performed the analyses. Andrey Ivanov manually gathered data about tool infrastructure. John Chamberlin gathered data about tool usage. David Hanauer, Brian Haas, Spyridon Bakas, Harry Hochheiser, Guilherme Del Fiol, Ava M. Hoffman, Richard L. Bradshaw, and Levi Waldron assisted Carrie Wright with the development of the survey.  All authors helped review the manuscript and were involved in discussions that contributed to the content and the overall direction of the data analysis, especially Mary J Goldman, Martin Morgan, Michael Reich, who presented about their experiences in software impact evaluation. Jeffrey T. Leek had the idea of evaluating possible associations of software usage with software infrastructure. David Hanauer, Candace Savonen, Mary J Goldman, Awan Afiaz, John Chamberlin, Alexander Getka, Aaron Holmes, Sarthak Pati, Dan Knight,  Spyridon Bakas, J. Gregory Caporaso, Patrick D. Schloss, Guilherme Del Fiol, Harry Hochheiser, and Carrie Wright contributed writing to the manuscript. 

\section{Acknowledgements}

This work was funded by the National Cancer Institute (NCI) of the National Institutes of Health (NIH), under award numbers UE5CA254170, U01CA242871, U24CA248265, and U54HG012517, as part of the Informatics Technology for Cancer Research (ITCR) program. The content of this publication is solely the responsibility of the authors and does not represent the official views of the NIH. We would also like to thank the survey participants who participated in our survey.

\bibliographystyle{unsrt} 

\bibliography{software_impact.bib}

\clearpage
\onecolumn

\rfoot{Afiaz et al.\hspace{7pt}$\mid$\hspace{7pt}2023\hspace{7pt}$\mid$\hspace{7pt}ar\textcolor{black}{X}$\chi$iv\hspace{7pt}$\mid$\hspace{7pt} Page S\thepage}

{\huge Supplemental Materials}

\hrule

\vspace*{0.5cm}

\begin{center}

{\Large Evaluation of software impact designed for biomedical research: Are we measuring what's meaningful?}

\vspace*{0.75cm}

{\large Awan Afiaz,  Andrey Ivanov, John Chamberlin, David Hanauer, Candace Savonen, Mary J Goldman, Martin Morgan, Michael Reich,  Alexander Getka, Aaron Holmes, Sarthak Pati, Dan Knight,  Paul C. Boutros, Spyridon Bakas, J. Gregory Caporaso,  Guilherme Del Fiol, Harry Hochheiser, Brian Haas, Patrick D. Schloss, James A. Eddy, Jake Albrecht, Andrey Fedorov, Levi Waldron, Ava M. Hoffman, Richard L. Bradshaw, Jeffrey T. Leek, Carrie Wright$^*$}

\vspace*{0.3cm}

{\small $^*$Correspondence to \url{cwright2@fredhutch.org}}

\end{center}

\setcounter{figure}{0}
\setcounter{table}{0}
\setcounter{section}{0}
\setcounter{page}{1}
\makeatletter
\renewcommand{\thetable}{S\@arabic\c@table}
\renewcommand{\theHtable}{S\@arabic\c@table}
\renewcommand{\thesection}{S\@arabic\c@section}
\renewcommand{\theHsection}{S\@arabic\c@section}

\vspace*{1cm}

{\bf \large Contents}

\begin{enumerate}
    \item \textbf{Supplemental Table~S1}
    \item \textbf{Supplemental Notes~S1-S5.}
\end{enumerate}

\clearpage

\noindent {\LARGE Supplemental Tables}

\begin{table}[!ht]
 \caption{\textbf{Survey Responses.} Responses to the question, "What would be your goals in evaluating the impact, engagement, or usage of a software tool?" Note that similar responses were considered under the same response category and individual pieces of responses were split to group similar elements of responses together.}
  \centering
  \begin{tabular}{|p{0.7\textwidth}|c|}
    \hline
    \rowcolor[gray]{.9} \textbf{Answer} & \textbf{Frequency} \\
    \hline
    Understand the extent to which a software tool is being used & 11 \\
    \hline
    Determining user needs and maximizing utility and impact for target users & 8 \\
    \hline
    Accrue information, devise metrics, and measure milestones to support grant reporting and funding & 8 \\
    \hline
    Evaluate if the software tool satisfies user requirements & 7 \\
    \hline
    Improve software and optimize development & 7 \\
    \hline
    Understand ease of use for new users, improve usability & 5 \\
    \hline
    Inform efforts, direction of research & 4 \\
    \hline
    Identifying popular features & 3 \\
    \hline
    Evaluating adaptability and integration with other tools & 3 \\
    \hline
    Improve community reach and engagement & 2 \\
    \hline
    Understand clinical significance, adoption by healthcare professionals, and effect on patient care & 2 \\
    \hline
    "Quantitative assessment and goal setting" and help decide where to focus resources & 2 \\
    \hline
    "Understand the value brought by the software and data." & 1 \\
    \hline
    "Identify ways to engage a broader community and increase userbase." & 1 \\
    \hline
    "Gather data to drive decisions around software practices." & 1 \\
    \hline
    "Measuring the usage of individual system components and understanding and planning for periodic fluctuations in demand." & 1 \\
    \hline
    Source of motivation to continue building the tool and improving the tool & 1 \\
    \hline
    Gather information to support publications & 1 \\
    \hline
  \end{tabular}
  \label{tab:survey_table}
\end{table}

\clearpage

\noindent {\LARGE Supplemental Notes}

\section{Methods}
\label{sec-supp-note-methods}

Greater detail, as well as information about data access, and code access about the survey and the manual tool characteristics evaluation can be found at this \href{https://hutchdatascience.org/ITCR_Metrics_manuscript_website/}{website}.

\subsection{Survey}
\label{sec-supp-note-methods-survey}
Forty eight participants in the ITCR program were asked a series of questions about their attitudes on evaluating the impact of scientific software. Some questions were more general, while others asked about practices for their most mature tool. These responses were collected anonymously, however respondents could include a link to their tool if they felt comfortable. The raw survey responses will not be made available to the public to protect the respondents. The survey questions can be found at this \href{https://hutchdatascience.org/ITCR_Metrics_manuscript_website/questionnaire.html}{webpage}.

\subsection{Manual Tool Characteristic Evaluation}

Documentation depth: The binary variable “Documentation depth” characterizes the degree of documentation for each tool as either “extensive” or “Webpage/PDF/ReadMe”. The term extensive documentation refers to comprehensive resources such as user guides, training, tutorials, code examples, and FAQs that allow users to better understand the functionality of the tools and interpret their outputs. The “Webpage/PDF/ReadMe” category, on the other hand, designates that the tool offers less information, possibly in the form of a website, PDF, or readme file. This "Webpage/PDF/ReadMe" category can be interpreted as having a lower level of documentation depth, as it typically provides users with a more basic level of information about the tool. This type of documentation may only include a brief description of the tool, its features, and how to install it without going into more detailed instructions or use cases. The "Webpage/PDF/ReadMe" category may be sufficient for users who are already familiar with similar tools or have prior experience in the field, but it may not provide enough guidance for beginners or users who need more detailed instructions to use the tool effectively. Therefore, having a higher level of documentation depth, such as "extensive,", can be beneficial for both novice and experienced users, as it provides more detailed information, tutorials, and use cases, making it easier for them to learn about the tool and utilize it to its full potential.

Instruction on How to cite: The binary variable "Instruction on How to cite" specifies whether each tool provides instructions on how to properly cite it in a research publication or other work. These instructions typically include information such as the author(s) names, version number, publication date, and website or repository where the tool can be found. Tools that do not provide such instructions can make it difficult for researchers to correctly cite the tool, potentially leading to incomplete or inaccurate acknowledgements of its use in research.

Twitter Presence: The variable "Twitter Presence" indicates whether a tool has a formal presence on Twitter. Only 3 tools had other forms of social media including YouTube (two tools - one with twitter and one without) and gitter (one tool with twitter). Tools with a twitter presence have an active account that distributes news, updates, and information on the development, features, and usage of the tool. These accounts can allow communication between tool creators and users while facilitating the tool's exposure and accessibility.

Extensive Contact: The binary variable "Extensive Contact Ways" indicates whether a tool offers multiple ways for users to contact developers for support or other inquiries, such as email addresses, discussion boards, contact forms, and more. Tools with several contact options may be more user-friendly as they enable researchers to seek support or report issues with the tool more conveniently and effectively. In contrast, tools with inadequate or no contact methods may be less attractive to researchers, as they don't provide timely help for technical difficulties or other concerns.

\subsection{Classification of software mention types}

Software mentions in scholarly articles can be categorized into four types \cite{schindler_annotation_2019, schindler_somesci-_2021}: allusion, usage, creation, and deposition. The distinction between these types can help researchers understand the role and availability of software in scholarly publications. We distinguish the aforementioned classifications of mention types as follows:
 
\textbf{Allusion:} This type of mention simply refers to the name of the software and does not require any indication of its usage. Allusions are used to state a fact about the software or to compare multiple software options for a problem. These are similar to the typical scholarly citations used to refer to related work.
 
\textbf{Usage:} A usage type mention occurs when software is used in an investigation and contributes to the study. This type of mention allows for conclusions about the study's origins and can be used to develop impact metrics. 
 
\textbf{Creation:} When new software is introduced in a publication, it is a creation type mention. This type of mention can be used to provide scholarly credit to the software's authors, as well as map new creations and track down original software publications.
 
\textbf{Deposition:} The publication of new software is regarded as a deposition mention. By including information about the publication, such as a license or URL, deposition expands on the creation type mention. When describing availability and licensing information for software, cross-references are used to annotate indirect statements about the software.
 
In summary, allusion-type mentions provide a broad overview of the presence of software in the text, while usage mentions allow for conclusions regarding its actual contribution. Creation mentions are used to identify newly created software, and deposition mentions provide details on its publication and availability.

Our analyses fully excluded the creation type mentions. We focused on the usage type mentions as this metric facilitates the evaluation of impact of the software tools better than the other mention types.

\subsection{Software Mention Analysis}

SoftwareKG-PMC is a database of text-mined software mentions published by \cite{Kruger_usage_20}. In principle, this resource enables direct identification and analysis of the usage of ITCR tools in published literature present on PubMedCentral Open Access Subset as of January 2021. Access to this data requires setting up the database locally, as described by the developers at this \href{https://urldefense.com/v3/__https://github.com/f-krueger/SoftwareKG-PMC-Analysis__;!!GuAItXPztq0!hBzPcpXZ3SCNtkVgVI9KgwxhTOgm6JY1gKzCzCFfer1uhtTUx-sWnQyzKdfDCT0yogPVN8-gs4KZy97R-JH4PN6VNw20GNQ$}{GitHub repository}. We downloaded the raw ntriple data from this \href{https://urldefense.com/v3/__https://zenodo.org/record/7400022__;!!GuAItXPztq0!hBzPcpXZ3SCNtkVgVI9KgwxhTOgm6JY1gKzCzCFfer1uhtTUx-sWnQyzKdfDCT0yogPVN8-gs4KZy97R-JH4PN6VRftmtmY$}{SoftwareKG-PMC} and loaded these into a Virtuoso database using the tenforce/virtuoso docker image available at this \href{https://urldefense.com/v3/__https://hub.docker.com/r/tenforce/virtuoso/__;!!GuAItXPztq0!hBzPcpXZ3SCNtkVgVI9KgwxhTOgm6JY1gKzCzCFfer1uhtTUx-sWnQyzKdfDCT0yogPVN8-gs4KZy97R-JH4PN6V0PP1Kqo$}{link}. We adjusted the virtuoso configuration file for a machine with 64Gb of RAM. After the database is instantiated, the docker container can be closed and resumed on demand.

We queried the database instance from R using code adapted from the SofwareKG-PMC-Analysis code notebook. The SPARQL library version 1.16 was installed manually from the CRAN archive at this \href{https://urldefense.com/v3/__https://cran.r-project.org/src/contrib/Archive/SPARQL/__;!!GuAItXPztq0!hBzPcpXZ3SCNtkVgVI9KgwxhTOgm6JY1gKzCzCFfer1uhtTUx-sWnQyzKdfDCT0yogPVN8-gs4KZy97R-JH4PN6VyojsQgQ$}{link} as the package is no longer maintained. We constructed a query function which returns all articles which mention a given keyword (i.e., software name) in a particular mention category (allusion, usage, creation, or deposition). Usage was used for most further analyses. This yielded 73095 article-level mentions of any type, corresponding to 60970 unique articles for 36 ITCR tool name keywords. 8 tools were not identified, possibly due to recency of release, complexity of the tool name, or gaps in the accuracy or completeness of SoftwareKG and/or PubMedCentral. We further note that results are contaminated to a varying degree by software name homonyms. This type of collision is a chief limitation of SoftwareKG-PMC, especially for tools with simple or generic names. The query was performed in a case-insensitive manner due to the tendency for authors to adjust the capitalization of software names, such as JBrowse vs Jbrowse. SoftwareKG-PMC does not reliably aggregate these variations.

\section{Survey Qualitative Responses}
\label{sec-supp-survey}

Supplementary Table \ref{tab:survey_table} shows the responses of several ITCR developers in response to a question about their goals in evaluating their scientific software. The survey participants expressed diverse perspectives, highlighting the wide range of applications for such analyses. The qualitative feedback to the open-ended question was compiled and organized into a tabular format for enhanced readability, providing a comprehensive overview of the developers' inputs about their goals and objectives. Note that direct quotes from the respondents are included with quotation marks.

\section{Software Infrastructure to Enable Evaluation}
\label{sec-supp-note-inf}

\subsection{Web presence}
Tools vary in terms of web presence. For example, some tools are web-based, some tools simply rely on a README file in a code repository, with simple installation information, while others provide extensive information and documentation on a separate website. Google Analytics can allow for more fine-grained tracking of how users are interacting with web-based tools or websites and can be helpful for later assessments, in addition to improving access, as one can also perform search engine optimization (SEO) to further encourage the visibility. For example, a developer could track what parts of the documentation users appear to read more frequently and pair this with information about common errors. For web-based tools, great detail can be determined about user interaction. Again Google Analytics can be helpful or for tools using a web-server manual log file inspection, or a service like Cronitor \cite{cronitor}  if the tool relies on a web-based server using cron \cite{cron_2009} job scheduling can also be informative.  It is however important to be mindful of user privacy regulations, see section \ref{sec:legal_ethics}.

\subsection{Citability}
Another component that can assist with evaluations is providing users with a method of citation, as well as information about how to cite the software, as some users may not be familiar with such practices. This is often done by publishing a manuscript about the software in a scientific journal and providing information about the publication on the software website and code repository. In addition, digital object identifiers (DOI), from publishing platforms like Zenodo \cite{zenodo}, can be used for other less conventionally cited materials, such as documentation, case studies, data, and the actual software itself.

Services like Altmetric \cite{noauthor_altmetric_2015}, allow for deeper analysis of engagement for anything with a DOI. They provide reports and badges that can be added to websites or manuscripts (depending on where they are published) that indicate how often a DOI is cited in multiple sources besides scientific articles, such as blogs, news articles, Wikipedia, social media, and more.  Individuals or organizations can use Altmetric to get such reports by simply searching for the DOI. These reports also have links to individual social media posts, citing articles, and more. In addition, Semantic Scholar  \cite{noauthor_semantic_nodate} provides reports that indicate where citations have occurred within scientific articles.

\subsection{Documentation}
Documenting software not only helps guide users but can also enable collection of useful metrics.  Documentation on websites can be tracked and provide detailed information about such usage. If a particular page has more traffic, it may indicate that the aspect of the tool discussed on that page is either particularly popular or particularly confusing. As with all metrics, pairing information together can aid interpretation. Providing different types of documentation is another step that assures confidence in software, and enables varying roles to succeed in using the software. For example, command-line usage is important for the users of software, while API documentation is important for developers who may want to extend the software.

\subsection{Communications} 
Providing mechanisms for users to communicate with one another and with the developers can provide another avenue for understanding usage and engagement.

\subsubsection{Feedback mechanisms}
One very helpful method of obtaining software usage metrics is to have users directly provide feedback. Providing such feedback mechanisms helps to identify software weaknesses. However, few users provide feedback and often feedback requires interpretation. Individuals are also unlikely to provide positive feedback. Nevertheless, user feedback mechanisms can be powerful for gaining insight from users. 
Feedback mechanisms can be passive or active in nature. More passive mechanisms may involve providing an email address on a website or simply allowing users to post an issue on a GitHub repository. More active mechanisms may include usability testing interviews, Google forms or other surveys, or providing automated GitHub issue templates to encourage specific kinds of feedback.

\subsubsection{Email and support forums}
While it’s possible to collect user feedback and provide support through email,  if a public support forum is used instead, users can learn from answers to questions of other users, reducing the burden on developers, and provide another opportunity for easier tracking of software engagement. This also provides an opportunity to build a community around the software and to learn what is and isn’t working for users. Some forum platforms support upvoting (where other forum users can upvote questions or feature requests). Using an existing support forum platform such as Discourse (\url{https://www.discourse.org/}) provides many features that are convenient in forum moderation, such as user trust levels, categorization and tagging of topics, and detailed tracking of forum posts, forum post views, and user and moderator activity levels that serve as useful metrics about a software project and the community around it.

Engaging with users about what new features they are excited about and acting on those discussions, is a great way to get informed feedback while rewarding user involvement. To further support community, developers could consider inviting user to attend workshops on new features or inviting them to help teach workshops on fundamentals.  Forum members who provide intellectual contributions to projects, including technical support, feature suggestions, highly referenced posts, etc, should be included as authors on relevant papers. And public facing activity summaries (e.g., \url{https://forum.qiime2.org/u/gregcaporaso/summary}) can also be useful for resume building for early career contributors.

Emailing newsletters to registered users, can be a useful method for informing users about new features, updates or issues. Systems like Mailchimp \cite{mailchimp} or HubSpot \cite{hubspot} can allow analytics about how often recipients open the news letter, click on links, or unsubscribe.

\subsection{Usability testing}

Usability refers to the ease of use for an individual to use software. Usability testing is a method of purposefully investigating the usability of a software tool. Usability testing can be highly powerful for collecting valuable metrics and information regarding the usage. This involves applying the scientific method to investigating a user's experience with a software tool and asking users to use the software tool while the tester observes their interactions with the tool. 

Often research institutions do not have these experts in user design and testing on staff. However, developers can and should utilize these techniques to conduct their own informal usability testing. Step-by-step guides of how to conduct usability testing have been published elsewhere \cite{savonen_2021}.  It should be noted that the benefits of usability testing are high even when only a few users are observed. 

\subsection{Workshops}
Hands-on workshops (online\cite{dillon_experiences_2021} or in-person) for your software are an excellent way to build a user community and get feedback. Workshops allow developers identify what confuses or challenges users. This can be illuminating as the challenges are often not what developers expect. Attendees can also be asked to participate in surveys about what they like and don't like to help guide future development. The quantity, duration, and attendance at your workshops are metrics that can be reported to funding agencies in grant proposals or reports. Posting recordings of events can be shared on YouTube or other platforms, which allows for other useful metrics. 

\subsection{Code of conduct}
Before engaging with user or developer communities, it is important to develop and publish a code of conduct for your project to outline expectations of behavior and how contributors can report violations. A code of conduct can be reassuring of community health and are required by some scientific software funders, such as the Chan Zuckerberg Initative’s Essential Open Source Software for Science program. Thus the presence of a Code of Conduct may be a measure that is assessed by potential funders. A good starting point for a code of conduct is the Contributor Covenant (https://www.contributor-covenant.org/), which can be adapted. Adopting and effectively managing\cite{aurora_how_2019} a  Code of Conduct can support the growth of your community by indicating a safe space, thereby encouraging engagement by individuals who might feel intimidated about getting publicly involved.

\subsection{Social media}
Having a social media presence through platforms such as twitter, instagram, and youtube, provides opportunities to track engagement with social media posts.  Pairing this with evaluations of engagement with the software itself, to determine if outreach strategies are successful. This can also be helpful to determine if documentation resources are useful. For example, the number of video views on a youtube documentation video can be informative to know what percentage of users may have actually seen the documentation. Videos with many views can also reassure others about using the software . 

\subsection{Reviews}
There are review mechanisms that can help reassure users about software. For example, SourceForge\cite{sourceforge} a platform for publishing and developing software allows users to rate software on the platform. Developers can integrate their GitHub repositories with SorceForge to take advantage of this review platform. Alternatively, GitHub also has a system of adding stars or followers to repositories, however, this appears to be somewhat inconsistently done in the community. 

\subsection{Version control mechanisms}
Many bioinformatics software projects maintain their source code under public revision or version control, using a platform such as GitHub or GitLab. Provided that the platform is actually used for version control (as opposed to hosting one or two versions of the software), this is a useful indicator of project health. Users can be confident that changes to the code are tracked to some extent and can also observe how much activity and updates have occurred to the software -- this can indicate to the user and developers of the project how recently maintenance has occurred. This can also help users identify the specific version or revision of the software they are using to record and report their methods, reproduce an analysis, or determine if they are impacted by a bug in the software. Furthermore, GitHub provides an insight tab with information about development metrics and its API allows for collection of these metrics in a systematic and reproducible way. 

\section{Software Health Infrastructure}
\label{sec-supp-note-health-inf}

\subsection{Development activity}
Finally project health can be assessed based on activity. By reviewing the frequency of commits and the date of the most recent commit in a revision control system, it is possible to assess whether the project is actively developed. Inactively developed projects project may not include fixes for recently discovered bugs, or improvements in methods to keep current with the field. Tracking who is making commits to the project can also provide details on whether the project is the work of one or a few developers, or of a large and active developer community. Neither a big or small developer community is necessarily better, though smaller developer communities can pose greater sustainability risk. If there is only one person who knows how to work on the code, and that person becomes unable or unwilling to continue working on the project, development or maintenance of the project may be discontinued.

\subsection{Automated test coverage and continuous integration testing}
Bugs in scientific software can have far-reaching impacts. Automated software tests allow developers to confirm that their code is behaving as expected under a range of normal and abnormal conditions. Systematically developed tests that can be run by both developers and users with a single command, give confidence that the software is doing what it is supposed to be doing. Platforms such as GitHub allow developers to advertise their test code coverage (the fraction of lines of code in the project that are covered by automated tests) and current build status (including whether tests are currently passing or failing). The presence or absence of this information, as well as quantitative metrics like test code coverage, are useful indicators of project health. Similarly, the use of automated or continuous integration testing, which can be configured to run when a change is committed to a software repository, when software releases are created, or at any other point where quality assurance is necessary to provide early signals of errors, is also an indicator of project health. This can help ensure that bugs or other issues can be detected before software is deployed by users.

The presence of different types of automated tests can be seen as an indicator of software maturity, and the care taken in the process of developing the software. Unit tests provide confidence in individual pieces of code; component and integration tests provide confidence that individual pieces of code behave correctly together; acceptance tests provide confidence that the overall software behaviors are correct. Achieving this in-depth test coverage requires careful software design upfront and throughout development. It is worth noting that, while useful, test coverage is not always a trustworthy representation of test quality. It is simply a measure of the proportion of code which is run by the test suite. It does not evaluate the quality of the test cases or assertions.

Other related static code analysis metrics can be used in conjunction with test coverage to build a more well-rounded picture of code quality. For example, cyclomatic complexity measures the logical complexity of a function. Code with high cyclomatic complexity contains numerous branching paths with variations in execution, and therefore should be avoided. A complex function can have 100\% test coverage without coming close to testing all possible outcomes. In fact, writing tests for highly complex code is often a lost cause. By considering cyclomatic complexity alongside test coverage, programmers can target specific areas of a codebase that need more tests. However, they can also see where they should instead prioritize refactoring, reducing complexity and improving testability first.

\subsection{Licensing}
In general, a software project should have a discoverable license that allows users to understand how they are and are not allowed to use the software. Licenses differ in how permissive they are, but the lack of a license defaults to highly restrictive terms and is an indicator that use of the software by others hasn’t been adequately considered. Similarly, the choice of license can be another indicator of inadequate usage consideration due to the restrictive and binding nature of some software licenses.

\section{Security}
\label{sec-supp-note-sec}

In any repository that contains source code, it is of vital importance that authentication secrets, such as API keys or cryptographic secrets be managed properly \cite{merkel_docker_nodate}. It becomes even more important in the case of code that deals with healthcare data, where even the slightest mishap in this regard could potentially expose protected health information (PHI) data and/or information to the public. During the continuous integration (CI) step of any project, incorporation of reputed static code analysis tools \cite{louridas_static_2006, ludwig_compiling_2017} (which are often available to open-source projects at no cost) ensures that tracking of the perpetually evolving landscape of security threats is offloaded to an entity that is specialized for its detection. It also allows the code to scale at a quick pace while ensuring bugs can be identified early in the development process, and finally allows software architects to define security guidelines to assist developers without impeding on their development process.
It is understood and accepted that open-source code is built “on the shoulders of others”, and regular analysis of the underlying dependencies (such as using the “Dependabot” product from GitHub \cite{noauthor_dependabot_nodate}) allows the automatic detection of security vulnerabilities (and potential mitigation strategies presented) to the development team in question \cite{alfadel_use_2021}. Products such as Dependabot maintain a frequently updated collection of packages with version/commit identifiers that contain security vulnerabilities with their fixes as defined by the broader community, and as such should be incorporated into all open-source repositories. Another set of security considerations come when packaging and deploying code in the form of containers, and more specifically using Dockers \cite{merkel_docker_nodate}. Depending on the veracity of the adversary in question, attacks on Docker containers can range from targeting the container itself, the host system, any collocated containers, the container management system or the source code itself\cite{combe_docker_2016}. One of the most common security threats from Docker containers come in the form of elevated user access \cite{combe_docker_2016, bui_analysis_2015}, which can be a critical flaw when running containers in an High Performance Computing (HPC) environment \cite{sparks_enabling_2019, bacis_dockerpolicymodules_2015}. This is specifically addressed using other containerization protocols such Singularity \cite{kurtzer_singularity_2017} and Podman \cite{gantikow_rootless_2020}. Though a single solution might not work for all cases, when deploying open-source software using the Docker containerization protocol, specific security mitigation strategies should be considered \cite{yasrab_mitigating_2021}.
\makeatother

\end{document}